\documentclass[aps,showkeys,showpacs,11pt,preprintnumbers,nofootinbib,superscriptaddress]{revtex4}
\usepackage[letterpaper]{geometry}                
\usepackage{graphicx}
\usepackage{amssymb,amsmath}
\usepackage{epstopdf}
\usepackage{yfonts}
\def\a{\alpha}		\def\b{\beta}		
\def\d{\delta}				
\def\g{\gamma}				
			\def\k{\kappa}		\def\l{\lambda}
\def\m{\mu}		\def\n{\nu}			
					\def\r{\rho}
					
					\def\D{\Delta}
		\def\G{\Gamma}

\DeclareMathOperator{\tr}{tr}
\def\be{\begin{equation}}
\def\ee{\end{equation}}

\begin{document}

\preprint{BRX-TH-613}
\preprint{CALT 68-2752}

\title{Gravity from self-interaction redux}

\author{S. Deser}
\email{deser@brandeis.edu}

\affiliation{Physics Department, Brandeis University, Waltham, MA 02454 and \\
 Lauritsen Laboratory, California Institute of Technology, Pasadena, CA 91125}

\date{\today}                                           

\begin{abstract}
 I correct some recent misunderstandings about, and amplify some details of, an old explicit non-geometrical derivation of GR. 
\end{abstract}

\maketitle
Long ago [1], I presented a compact derivation of GR from  an initial free flat space long-range symmetric spin two field: Since special relativity replaces the matter Newtonian scalar mass density by its stress-tensor, a tensor must likewise replace the scalar "potential". Consistency then forces this field to couple to its its own stress tensor if it is to allow any matter coupling: it either stays free-- and dull-- or its stress- tensor must be added to that of matter as the field's source. This bootstrap was then explicitly performed in GR by exploiting its first derivative, cubic, $L \sim p\dot{q}- q p^2$, rather than its more familiar second-order non-polynomial $L(q)$, form. The process was also illustrated in the simpler, but precisely analogous, context of deriving (nonlinear)YM from a multiplet of free Maxwell fields, which must likewise self-couple to accept non-abelian sources. Subsequently, two extensions of [1] were found: First, it was generalized to allow starting from any constant curvature background, where spin 2 is consistently defined [2]. The cosmological term could then also be included in the bootstrap. Second, a tree-level quantum derivation [3] (later generalized to include SUGRA [4]) provided an alternate framework, where the irrelevance of inherent field redefinition ambiguities and freedoms is particularly clear.  

Recently, however, there have appeared lengthy, (if not mutually consistent) critiques [5,6] of [1]. This note addresses and resolves their concerns, both conceptual and technical, by expanding on the, perhaps too concise, original. For orientation, we start with the list of main worries and the short answers.

1. The self-coupling idea, while appealing, does not work out concretely; also,the gravitational stress-tensor is ill-defined. 

These worries stem from too narrow a view of self-coupling and a too broad one of non-uniqueness.  Self-coupling means  that the right-hand side of the original free-field equations, in one of its possible incarnations, acquires as a source the field's own total stress tensor. This will be (re-)derived below,  using the equivalent but more convenient Ricci, rather than Einstein, form of the equations. A related complaint was that the coupling did not appear in the naive, $h^{\m\n} T_{\m\n}$, form in the action. True, but irrelevant: to repeat, the only physical requirement is that, in the field equations, the full $T_{\m\n}$ become the source of the originally free field; the action's sole job is to yield these, and it does --- see (13) below.  Non-uniqueness of the stress tensor: it is indeed always undetermined up to identically conserved super-potentials. Further, while the one place where this non-uniqueness is relevant, namely when the stress tensors become local sources, is here, it is also precisely here that all such ambiguities can be absorbed, as we shall see, by harmless field redefinitions. Another non-uniqueness pseudo-problem is that free gauge fields of spin $>1$ cannot possess (abelian) gauge-invariant stress tensors; this truism actually turns out to be a plus: only full GR recaptures the initial invariance, but now in non-abelian form, at the (satisfactory!) price of forfeiting any physical significance for its own stress-tensors, a fact also known as the equivalence principle. The only   restriction on the initial stress-tensor(s) is that they be symmetric so they can drive the graviton's symmetric field equations; further, only they can define angular momentum.   

2. The GR action's non-analytic dependence on the Einstein constant $\k$ cannot be obtained perturbatively starting from the, $\sim \k^0$, free field. 

This worry will be easily dispatched in its place; simply, the final $1/\k^2$ dependence arises from a constant field rescaling of the (analytic) result to connect the field theoretical and geometrical variables' dimensions.

3. The theory's second derivative order was an assumption.
  
This is as true here as it was for Einstein and Newton! Formally, GR is but one of an infinite set of geometrical models, with as as many derivatives as desired (e.g., $L \sim R D^n R$)...Observation determines the initial kinematics, excluding (to leading order at least) scalar-tensor mixtures and higher derivative terms. Most relevant for us, second derivative order together with infinite range (any finite range makes qualitatively wrong weak-field predictions [7]) means
that a gauge invariant (i.e., ghost-free) massless tensor field is the initial, special relativistic, mediator of matter-matter forces (their attractive sign then being a built-in bonus [8]). 

4. Total divergences and surface terms are important.  

Yes, but not to obtain Euler-Lagrange equations from an action. Surface terms are indeed physically useful in GR, but not because of their presence in its action, contrary to myth. 

5. As (correctly) noted in [5], there have many other attempts at deriving GR from self-coupling, none of which succeeded: their approach being purely metric, the infinite summations needed to reach non-polynomial metric GR have never been performed. Instead, they were replaced by such statements as "what else could it sum to?" and "the sum must be general covariant, ergo GR". 

Agreed. In particular the covariance of the final result, in the strong sense of being achieved without involving an external metric, does emerge here without being postulated; likewise, "summation" is trivial.
 
For maximum clarity, we focus on the logic, with a minimum of formalism and indices; that can be found in [1]. The flat space, first order, Fierz-Pauli massless spin 2 Lagrangian is
\begin{equation}
L_2 = h^{\m\n} (\partial_\a \Gamma^\a_{\m\n} - \partial_\m \Gamma^\a_{\a\n}) +\eta^{\m\n}( \Gamma^\a_{\m\n} \Gamma^\b_{\b\a}-\Gamma^\a_{\b\m}\Gamma^\b_{\a\n})
\end{equation}
The two independent variables are the Minkowski tensors ($h^{\m\n}$,  $\G^\a_{\m\n}$), with dimension ($L^{-1}$, $L^{-2}$) as befits their "$(q,p)$" nature; $\eta$ is the Minkowski metric.  The resulting first order field equations
 \begin{eqnarray}
\partial_\a \Gamma^\a_{\m\n} - {1\over 2}(\partial_\m \G^\a_{\a\n} + \partial_\n \G^\a_{\a\m}) &=& 0  \\\partial_\a h^{\m\n} - \partial_\m h^{\n\a}-{1\over 2} \eta_{\m\n} \partial_\a h^\b{}_\b &=& 2 \G^\a_{\m\n} - \eta^\a_\m \G^\b_{\b\n} - \eta^\a_\n \G^\b_{\b\m}
\end{eqnarray}
are equivalent to
\begin{equation}
2 R^L_{\m\n}(h) \equiv \partial_\b \partial^\b( h^{\m\n} -{1\over2} \eta_{\m\n} h^\a_\a) - \partial_\n \partial_\a h^{\m\a} -   \partial_\m \partial_\a h^{\n\a}
\end{equation}
in terms of the linearized Ricci (rather than Einstein) tensor.\footnote{For comparison, the first order vector theory equivalents are the initial, $L_1 \sim F curl A-F^2$ and $L_{YM} \sim L_1+ gFAA \sim p\dot{q} -p^2+ pq^2$ as final, forms; they are spelled out in [1].} [Our $h^{\m\n}$ is related to the usual covariant metric deviation $h_{\m\n}$ by $h^{\m\n}= -h_{\m\n}+(1/2) \eta_{\m\n}(h_{\a\b}\eta^{\a\b})$.]  Note however that our $h^{\m\n}$ is {\it not} the start of an expansion, but is the total deviation, from its Minkowski value, of the full contravariant metric density.

The full GR, Palatini, Lagrangian we want to derive is 
\begin{equation}
L_{EH}({\cal{G}}, \G)= {\k^{-2}} {\cal{G}}^{\m\n} R_{\m\n}(\G) = {\k^{-2}} {\cal{G}}^{\m\n} \left(  \partial_\a\Gamma^\a_{\m\n} - \partial_\m \Gamma^\a_{\a\n} +\Gamma^\a_{\m\n} \Gamma^\b_{\b\a}-\Gamma^\a_{\b\m}\Gamma^\b_{\a\n}\right); 
\end{equation}
$\cal{G}$ is the contravariant metric density, $\Gamma$ the (independent) affinity. The chief differences between (1) and (5) are that there is no background space dependence in (5), and that it is cubic (rather than quadratic) in the fields. This latter property is its compelling attraction for us, in contrast to the  second order metric formulation's non-polynomial dependence on both the metric and its inverse through the affinity's metric dependence. The GR equations, from varying ${\cal{G}}$ and $\G$ independently, are 
\begin{eqnarray}
R_{\m\n}(\Gamma) \equiv \partial_\a \Gamma^\a_{\m\n} - {1\over2} \partial_\m \Gamma^\a_{\a\n}-{1\over2} \partial_\n \Gamma^\a_{\a\m} +( \Gamma^\a_{\m\n} \Gamma^\b_{\b\a}-\Gamma^\a_{\b\m}\Gamma^\b_{\a\n}) &=& 0, \\
- \partial_\a {\cal{G}}^{\m\n} + {\cal{G}}^{\m\n} \G^\l_{\l\a} - {\cal{G}}^{\m\r} \G^\n_{\a\r}-{\cal{G}}^{\n\r} \G^\m_{\a\r} &=& 0,      
\end{eqnarray}
and reduce to $R_{\m\n}({\cal{G}}) = 0$ upon inserting $\G({\cal{G}}) \sim {\cal{G}}^{-1} \partial {\cal{G}}$ into (6). Note that the geometrical variables' dimensions are (${\cal{G}} \sim L^0$, $\G \sim L^{-1}$). We will see that the non-analyticity of (5) is purely apparent, being removable by constant rescalings. It is useful for the sequel to express this desired answer in flat space notation by expanding (5) in terms of ${\cal{G}}=\eta+\k h$ ($\k$ restores $h$'s original dimension $L^{-1}$) and to restore its old dimension to $\G$, by defining $\overline{\G}=\k^{-1} \G$; we now drop all indices to concentrate on the form and logic:
\begin{equation}
L_{EH}(h, \overline{\Gamma})= \k^{-1} \eta \partial \overline{\Gamma} + (h \partial \overline{\Gamma}+ \eta \overline{\Gamma}~\overline{\Gamma}) +\k h \overline{\Gamma}~ \overline{\Gamma}.
\end{equation}
The first term being an irrelevant total divergence, $\k$ now appears quite tamely in the rest of (8), disposing nicely of that worry. The middle terms are precisely the quadratic free field Lagrangian (1). The cubic term, $\k h \overline{\G}\overline{\G} \equiv \k h S$ is of course supposed to supply the heralded self-coupling of $h$ to its stress tensor in the field equations (as we will check it does), the very reason S is not itself the stress tensor. Given this flat space form of GR, it remains to show that the cubic term in (8) is the right choice: does it provide just the right (whatever that is) stress tensor source of the free field--middle terms'--field equation? The justification has three parts: first obtaining the stress tensor(s) of the middle terms' action $\cal{I}$, then showing why its non-uniqueness (including abelian gauge-variance) is harmless, and finally verifying that the chosen cubic term (the one that agrees with $L_{EH}$) indeed produces this stress tensor. 

First, the stress tensor: We use the Belinfante prescription: write the flat space action covariantly with respect to a fictitious auxiliary metric (for us a contravariant density) $\gamma^{\m\n}$, vary the resulting action with respect to it, then set it back to $\eta^{\m\n}$ in the resulting variation. The result is a symmetric on-shell, trace-shifted, stress tensor. In (1), there are two places to covariantize: the obvious $\eta \Gamma \Gamma \rightarrow \gamma \Gamma \Gamma$ and $h \partial \G \rightarrow h D(\gamma)\G$, where $D$ is the covariant tensor derivative involving the auxiliary Christoffel symbols $\sim (\partial \g)$ to first order.  Manifestly,
\begin{equation}
\overline{T}_{\m\n} \equiv T_{\m\n}-{1 \over 2} \eta_{\m\n} \tr T \equiv \left({\delta \cal{I} / \delta \gamma} \right)|_{\gamma=\eta} \sim  \partial (h\G)+\G\G.   
\end{equation}
Next (non-)uniqueness: to the Belinfante tensor (of any system) may be added any identically conserved super-potential
\begin{equation}
\D_{\m\n} = \partial^{\a}\partial^{\b} H_{[\m \a][\n \b]}=\D_{\n\m}, ~~~ \partial_\m \D^{\m\n} \equiv 0, 
\end{equation}
where $H$ is any 4-index function with the symmetries of the Riemann tensor, to keep $\D$ symmetric. [These contributions may also be thought of as the result of adding non-minimal couplings $\sim R_{\a\b\g\d} (\gamma) F^{\a\b\g\d}(h, \G)$ to the original action (before varying $\gamma$).] But identical conservation of $\D$ means precisely that it can be absorbed by field redefinition: the usual linearized Einstein equation is of the form 
\begin{equation}
G^L_{\m\n}(h)= {\cal{O}}_{\m\n \a\b} h^{\a\b}, ~~~   \partial^\m {\cal{O}}_{\m\n\a\b}  \equiv 0. 
\end{equation}
Hence any identically conserved source can simply be removed by a corresponding shift in $h$. [The initial Belinfante part, not being a super-potential, cannot be shifted away.] Finally, we must show that the cubic term in (8) indeed yields the desired field equation, with the stress tensor (9) as source of the free field. That is, we want to verify that the  full field equation reads $R^L_{\m\n}(\G(h)) \sim \k \overline{T}_{\m\n}$. The Einstein equations (6,7) are, dropping the overbars and expanding ${\cal{G}}$,
\begin{equation}
\partial \G +\k \G \G=0, ~~~   \G= \partial h+ \k h\G.                   
\end{equation}
Differentiating the second and inserting it into the first equation gives precisely the promised second order form 
\begin{equation}
\partial^2 h=\k [\partial(h\G)+ \G\G] \equiv \kappa \bar{T}.
\end{equation}
More explicitly, the left side is $R^L _{\m\n} (\G(h))$, while the right is just the $\overline{T}_{\m\n}$ of (9) if (and only if) we use the cubic term of the GR action (5). Equally important, the bootstrap stops here because this cubic term in the action does not generate any further (cubic) stress-tensor correction, being both $\eta$-and derivative-independent. This completes our exegesis.  

Sources: it is rather obvious that any matter action must couple to the final GR through its variables $({\cal{G}}, \G)$ or ${\cal{G}}$ alone, and do so covariantly in order to respect the GR equation's Bianchi identities  by having an (on-shell) covariantly conserved metric variation. But this is just Noether's theorem: any system's stress-tensor, namely the variation of its action with respect to the metric that makes it invariant, is covariantly conserved by virtue of its own field equations, irrespective of the equations (if any), satisfied by the metric.

In summary, I have annotated the steps involved in the non-geometric derivation [1] of GR from special relativistic field theory as the unique consistent self-interacting system, (13) extending the initial free massless spin 2. The main ingredients were: computing the field's standard Belinfante stress tensor, invoking field-redefinition freedom to neutralize its non-uniqueness, performing a constant field rescaling to relate geometric and field theoretic variables, and (most important) employing the cubic, Palatini, first order forms to permit explicit, trivial, summation. It goes without saying that this non-geometrical interpretation of GR,
far from replacing Einstein's original geometrical vision, is a tribute to its scope.
 
This work was supported by NSF grant PHY 07-57190 and DOE grant DE-FG02-92ER40701                         

\textbf{References:}

1. S.Deser, Gen Rel Grav 1 9(1970), reprinted as gr-qc/0411023.

2. S.Deser, Class Quantum Grav 4 L99(1987).

3. D.Boulware and S.Deser,  Ann Phys 89 193(1975).

4. D.Boulware, S.Deser and J.Kay, Physica 96A 141(1979).    

5. T.Padmanabhan, Int J Mod Phys D17 367(2008), gr-qc/0409089.

6. L.Butcher, M.Hobson and A.Lasenby, Phys. Rev D80 084014(2009), gr-qc/0906.0926.

7. H.van Dam and M.Veltman, Nucl. Phys. B22 397(1970);  V.Zakharov, JETP Lett. 12, 312(1970); L.Faddeev and A.Slavnov, Theor.Math.Phys. 3 18(1970);  S.Wong, Phys. Rev. D3, 945(1971); 
I.Kogan, S.Mouslopoulos and A.Papazoglou, Phys.Lett.B503 173(2001), hep-th/0011138;  M.Porrati, Phys. Lett. B498 92(2001), hep-th/0011152.

8. S Deser, Am. J Phys. 73 6(2005), gr-qc/0411026.

\end{document}